\documentstyle[sprocl]{article}
\def\Journal#1#2#3#4{{#1} {\bf #2}, #3 (#4)}

\def\al{\alpha}

\def\be{\begin{equation}}
\def\ee{\end{equation}}
\def\bea{\begin{eqnarray}}
\def\eea{\end{eqnarray}}
\def\T{\textstyle}
\def\D{\displaystyle}
\begin{document}

\title{MUCH ADO ABOUT NOTHING: VACUUM AND RENORMALIZATION ON
THE LIGHT-FRONT}
\footnote{Lecture notes, based on three lectures given at the 
NUclear Summer School NUSS 97, at Seoul National University,
Seoul, South Korea, May 1997.}
\author{ Matthias Burkardt  }

\address{Department of Physics, New Mexico State University,\\ Las Cruces, NM 
88003-0001, USA}


\maketitle%
\abstracts{In the first part of my lectures, I will use the example 
of deep-inelastic scattering to explain why light-front coordinates 
play a distinguished role in many high energy scattering experiments.
After a brief introduction into the concept of light-front
quantization, I will show that the vacuum for any light-front
Hamiltonian is trivial, i.e. the same as for non-interacting fields.
In the rest of my lectures, I will discuss several toy models in
1+1 and 3+1 dimensions and discuss how effective light-front
Hamiltonians resolve the apparent paradox that results from having a 
trivial light-front vacuum. 
}
  
\section{The Uses of Light-Front Quantization}
\footnote{This Section has been taken from Ref. \cite{mb:adv}.}
In deep-inelastic scattering (DIS) experiments one scatters leptons
at very high energy from a nucleon or a nucleus. Typically, the
target nucleon gets destroyed in such a process and one obtains a
large number of particles in the final state. In the most simple
version of these experiments (inclusive), one completely ignores
all the details of the hadronic final state and only measures the
energy transfer $\nu = E-E^\prime$
and momentum transfer $Q^2=-q^2$ from the lepton on the target.
The double differential (inclusive) cross section for the scattering
angle and energy of the lepton can be written in the form
\begin{equation}
\frac{d^2\sigma}{d\Omega dE'} = \frac{4 \pi \alpha^2}{M Q^4} \;
\left\{ W_2(Q^2,\nu)\; \cos^2\,\frac{\theta}{2} +
        2 W_1(Q^2,\nu)\; \sin^2\,\frac{\theta}{2} \right\}.
\label{eq:dis}
\end{equation}
In Eq. (\ref{eq:dis}) the Rutherford cross section has been factored
out since it represents what one obtains for scattering from 
point-like targets. \footnote{For a point-like target, the
structure functions are just $\delta$ functions due to energy
conservation.}
In this expression, all details from the microscopic structure of the
target are parameterized in the structure functions $W_1$ and $W_2$.
In general, i.e. for arbitrary momentum transfer, these structure 
functions are functions are independent functions of two variables,
$\nu$ and $Q^2$. However, in the Bjorken limit, i.e. when
$Q^2\rightarrow \infty$ and $\nu \rightarrow \infty$, such that
$x\equiv Q^2/2M\nu $ fixed, it turns out that to a good approximation
$W_1$ depend only on the ratio $Q^2/2M\nu$
\begin{eqnarray}
2MW_1(Q^2,\nu)&\stackrel{Bj}{\longrightarrow}& F_1(x) 
\nonumber\\
\nu W_2(Q^2,\nu)&\stackrel{Bj}{\longrightarrow}& F_2(x) . 
\end{eqnarray}
\begin{figure}
\unitlength1.cm
\begin{picture}(14,4)(-1,-5.5)
\includegraphics{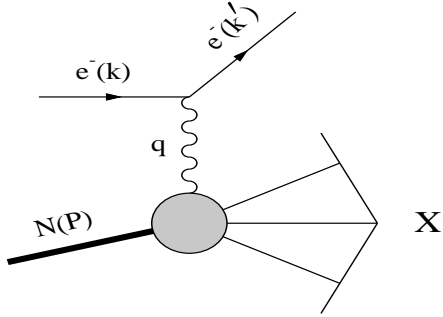}
\end{picture} 
\label{fig:dis}
\caption{DIS from a nucleon or nucleus. Only the scattered lepton
is measured.}
\end{figure}
This scaling behavior was the first direct evidence for point-like
charged constituents inside the proton, the quarks, and in the
parton model one describes hadrons as an ensemble of free quarks
and gluons, parameterized by the parton distribution functions
which are essentially the Bjorken-scaled structure functions.

These parton distribution functions contain an immense wealth of 
information about the quark and gluon structure of nucleons and 
nuclei. However, before one can relate this information to
phenomenological models or theoretical calculations, it is essential
to understand what these parton distributions really are.

For this purpose, we first use the optical theorem to rewrite the 
inclusive lepton-nucleon cross section in terms of the imaginary
part of the forward Compton amplitude 
\begin{figure}
\unitlength1.cm
\begin{picture}(14,6)(6,-11.)
\includegraphics{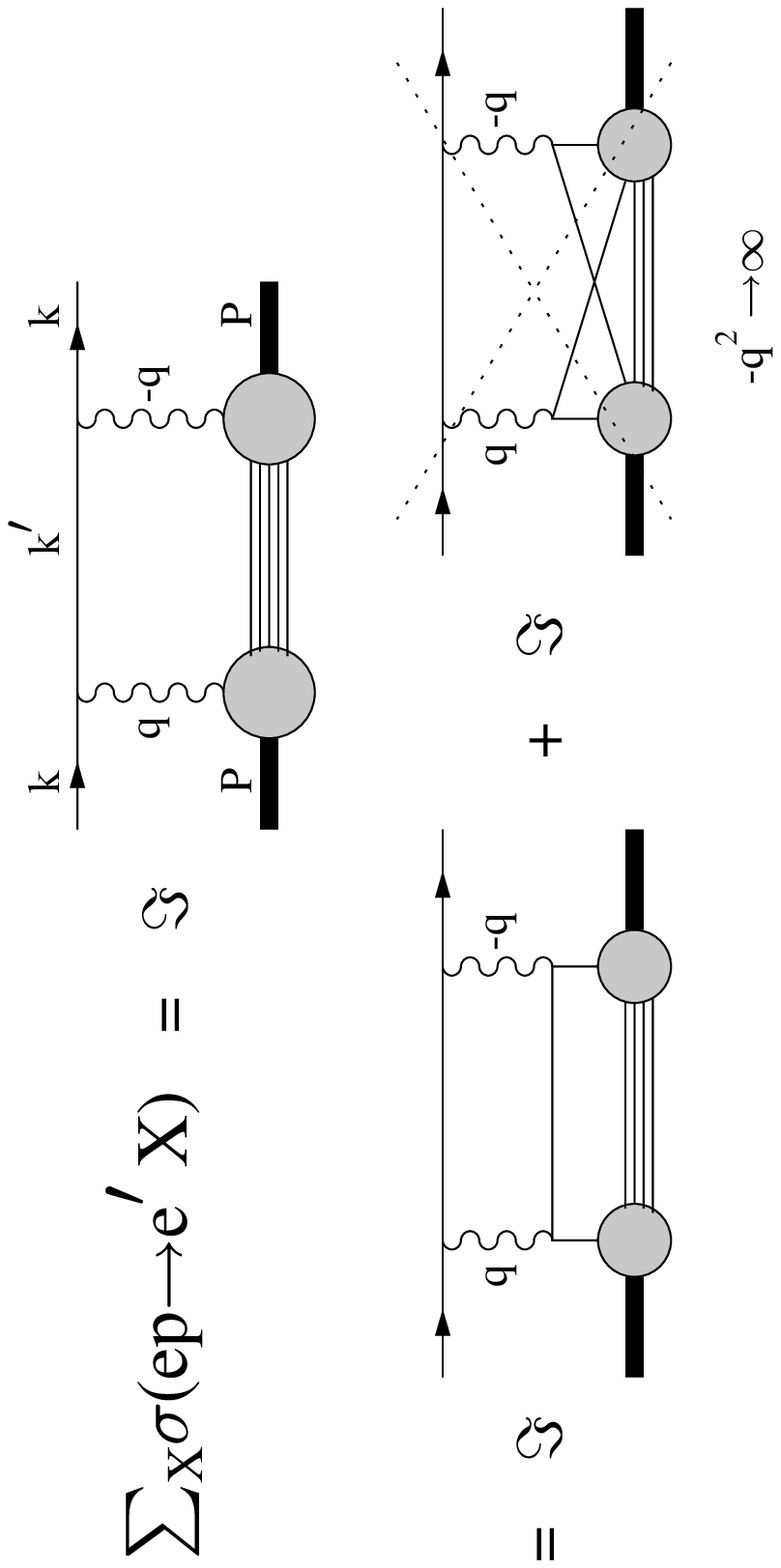}
\end{picture}
\label{fig:compt}
\caption{Using the optical theorem to relate the inclusive electron 
nucleon cross section to the imaginary part of the forward Compton
amplitude.}
\end{figure}
(Fig. 2).%\ref{fig:compt}).
In general, i.e. when $Q^2$ is not large, the two photons in Fig. 2
%\ref{fig:compt} 
can couple either to the same quark or to different
quarks. However, in the Bjorken limit, it is very difficult for the
transferred momentum  to flow from the first quark-photon vertex
to the second --- unless the two photons couple to the same quark.
If the two photons couple to different quarks then one needs additional
gluons to exchange the momentum and for this one has to ``pay'' with
additional energy denominators. Therefore, for large $Q^2$, the
crossed diagrams can be omitted.

In addition, since QCD is asymptotically free (i.e. the strong
coupling constant $\al_s(Q^2)$ vanishes for $Q^2\rightarrow \infty$),
one can neglect all interactions of the struck quark when it propagates
between the two quark-photon vertices.
Finally, since the struck quark is ultra-relativistic (and on its
mass shell since we take the imaginary part!) it propagates
almost along a light-like direction $x^\mu x_\mu =0$.

In summary, since the quark does not interact and moves along a 
light-like direction between the two quark-photon vertices,
it is as if one removes the quark at a certain point from the
proton and puts it back in at another point which is displaced
from the first vertex by a light-like distance.
It should therefore not be surprising that the parton distributions
measured in DIS are related to
correlation functions along a light-like direction:
\be
q(x) = \int \frac{d\zeta^-}{2\pi} \;
\langle P|\overline{q}(0) \gamma^+ q(\zeta^-)|P\rangle \;e^{i\zeta^-xP^+}
\label{eq:cor} .
\ee

It is instructive to compare DIS in non-relativistic systems
(e.g. in solid state physics) to the relativistic DIS which we
discussed above. The essential difference is that in non-relativistic
systems, if the momentum transfer is much larger than the intrinsic
momentum scale of the constituents in the target, the stuck particle
moves much faster than the spectators. As a result, its motion
can be approximated as infinitely fast, i.e. instantaneous and
one finds that DIS in non-relativistic systems probes
equal time correlation functions: ordinary momentum densities.  
In contrast, in relativistic systems where both the struck particle 
and the ``spectators'' have velocities that are of the order of the speed of light and
therefore, the motion cannot be approximated as instantaneous and
one probes correlations in a light-like direction: light-front
momentum densities.

This means that in a conventional framework, i.e.
when $x^0$ is time, parton distributions are related to real time
response functions. In order to calculate them it is not sufficient
to know the ground state wave function of the target, but one also
needs to know the couplings to excited states as well as the time
evolution of the excited states between the two couplings to external
photons. This makes it not only difficult to calculate parton 
distributions but also to interpret them.

The situation changes immediately when one introduces a new time
variable $x^+ \equiv
x^0+x^3$ (where $x^3$ is the direction of the momentum transfer).
With this choice of time, light-like correlation function
(more precisely: correlation functions in the $x^-=x^0-x^3$ direction)
are equal ``time'' correlation functions.
First of all, it is much easier to calculate in this framework
since parton distribution functions are simply momentum densities and
it is sufficient to know the ground state LF-wave function in order
to compute them. More importantly, since parton distributions are
such simple observables, it is much easier to develop an
intuitive understanding of otherwise obscure results.

There are many other examples of high energy scattering experiments
where LF coordinates also play a distinguished role, such as
certain exclusive reactions at large momentum transfer. Some of these
will be covered in Stan Brodsky's lecture notes. 

If one wants to take advantage of these results, it is necessary
to formulate QCD in a Hamiltonian framework where $x^+$ is time.
What this means will be discussed in Section \ref{sec:canoni}.
Quantizing on equal $x^+$ hypersurfaces is not a new idea.
The first work in this direction was by Dirac in 1949 \cite{dirac}.

LF quantization is very similar to canonical equal 
time (ET) quantization. Both are Hamiltonian formulations of
field theory, where one specifies the fields on a
particular initial surface. The evolution of the fields
off the initial surface is determined by the 
Lagrangian equations of motion. The main difference 
is the choice of the initial surface, $x^0=0$ for
ET and $x^+=0$ for the LF respectively. 
In both frameworks states are expanded in terms of fields
(and their derivatives) on this surface. Therefore,
the same physical state may have very different
wavefunctions\footnote{By ``wavefunction'' we mean here
the collection of all Fock space amplitudes.}
in the ET and LF approaches because fields at $x^0=0$ 
provide a different basis for expanding a state than 
fields at $x^+=0$. The reason is that the microscopic 
degrees of freedom --- field amplitudes at $x^0=0$ 
versus field amplitudes at $x^+=0$ --- are in general
quite different from each other in the two formalisms. 

This difference in the choice of microscopic degrees of
freedom is what explains how it is possible that a given
physical observable (e.g. parton distributions) may
be a very complicated object in the ET framework, and
at the same time be a relatively simple quantity in the
LF framework. 
\section{Canonical Quantization in Light-Front Coordinates}
\label{sec:canoni}
In this chapter, the formal steps
for quantization on the light-front are presented. 
\footnote{For a more detailed discussion, see the recent review in Ref.
\cite{big}.}
For pedagogical
reasons this will be done by comparing with conventional quantization
(with $x^0$ as ``time''). On the one hand this shows that the basic
steps in the quantization procedure in LF and in ET 
formalism are in fact very similar. More importantly, however,
we will thus be able to
highlight the essential differences between these two approaches
to quantum field theory more easily.

In the context of canonical quantization one usually starts from 
the action 
\begin{equation}
S = \int d^4x {\cal L}.
\end{equation}
(${\cal L}={\cal L}(\phi, \partial_\mu \phi)$)
After selecting a time direction
$\tau$ \footnote{Here $\tau$ may stand for ordinary time $x^0$ as well as
for LF time $x^+=\left(x^0+x^3\right)/\sqrt{2}$ or any other 
(not space-like) direction.}
one forms the momenta which are canonically conjugate to
$\phi$
\begin{equation}
\Pi (x)= \frac{ \delta {\cal L}}{\delta \partial_\tau \phi}
\end{equation}
and postulates canonical commutation relations 
between fields and corresponding momenta at equal ``time'' $\tau$
(Table \ref{tab:can}).
\footnote{The canonical quantization procedure in the ET
formulation can for example be found in Ref. \cite{bj:rel}.
The  rules for canonical
LF-quantization have been taken from Refs. \cite{yan}.}
%\newpage
\begin{table}
\begin{tabular}{c|c}
normal coordinates & light-front \\[1.5ex]
\hline
\multicolumn{2}{c}{coordinates}\\[1.5ex]
$\begin{array}{ll}
x^0 & \mbox{time} \\ x^1,x^2,x^3 & \mbox{space}
\end{array} $ &
$ \begin{array}{ll}
x^+ = \frac{\T x^0+x^3}{\T \sqrt{2}} & \mbox{time} \\
x^- = \frac{\T x^0-x^3}{\T \sqrt{2}}, x^1, x^2  & \mbox{space}
\end{array} $ \\[1.5ex]
\multicolumn{2}{c}{scalar product}\\[1.5ex]
$a \cdot b = a^0b^0-{\vec a}{\vec b} $ & 
$a \cdot b = a^+b^-+a^-b^+-{\vec a}_\perp $ \\
$=a^0b^0-a^1b^1-a^2b^2-a^3b^3 $ & $=a^+b^-+a^-b^+-a^1b^1-a^2b^2 $ \\[1.5ex]
\multicolumn{2}{c}{Lagrangian density}\\[1.5ex]
${\cal L} = \frac{1}{2} \left(\partial_0 \phi\right)^2
-\frac{1}{2}\left(
\stackrel{\rightarrow}{\nabla} \phi \right)^2 -V(\phi)
$ & ${\cal L} = \partial_+\phi \partial_-\phi
-\frac{1}{2}\left(
\stackrel{\rightarrow}{\nabla}_\perp \phi \right)^2 -V(\phi) $ \\[1.5ex]
\multicolumn{2}{c}{conjugate momenta}\\[1.5ex]
$\pi = \frac{\T \delta{\cal L}}{\T\delta\partial_0\varphi} =
 \partial_0\varphi$ &
$\pi = \frac{\T \delta{\cal L}}{\T\delta\partial_+\varphi} =
 \partial_-\varphi$ \\[1.5ex]
\multicolumn{2}{c}{canonical commutation relations}\\[1.5ex]
$ [\pi(\vec{x},t),\varphi(\vec{y},t)] $ &
$ [\pi(x^-,x_{\perp},x^+), \varphi(y^-,y_{\perp},x^+)] $ \\
$ = -i\delta^3(\vec{x}-\vec{y}) $ &
$ = -\frac{i}{2} \delta(x^- -y^-)
\delta^2({\vec x}_{\perp}-{\vec y}_{\perp}) $ 
%\footnote{Please note the factor $\frac{1}{2}$ in the LF commutator.
%It arises because the equation relating the momenta to the fields
%above contains no time derivative, i.e. it is a constraint equation.
%For a more detailed discussion and references, the reader is referred
%to the appendix.}
\\[1.5ex]
\multicolumn{2}{c}{Hamilton operator}\\[1.5ex]
$P^0 = {\D\int} d^3x\; \cal{H}(\varphi,\pi) $ &
$P_+ = {\D\int} dx^- {\D\int} d^2x_{\perp}\; {\cal H}(\varphi,\pi) $ \\
${\cal H} = \pi \partial_0 \varphi - {\cal L} $ &
${\cal H} = \pi\partial_+\varphi - {\cal L} $ \\[1.5ex]
\multicolumn{2}{c}{momentum operator}\\[1.5ex]
$\vec{P} = {\D\int} d^3x\; \pi \vec{\bigtriangledown}\varphi $ &
$P_- = {\D\int} dx^-d^2x_{\perp}\; \pi \partial_-\varphi $ \\
& ${\vec P}_\perp = {\D\int} dx^-d^2x_{\perp}\; 
\pi {\vec \partial}_{\perp}\varphi $ \\
\multicolumn{2}{c}{eigenvalue equation}\\[1.5ex]
$P^0 |\psi_n\rangle = E_n |\psi_n\rangle $ &
$P_+ |\psi_n\rangle = P_{+n} |\psi_n\rangle $ \\[1.5ex]
$\vec{P}$ fixed & $P_-, {\vec P}_{\perp}$ fixed \\[1.5ex]
\multicolumn{2}{c}{hadron masses} \\[1.5ex]
$M_n^2 = E_n^2 - \vec{P}^2 $ &
$M_n^2 = 2 P_{+n}P_- - {\vec P}_{\perp}^2 $
\end{tabular}
\caption{canonical quantization in ordinary coordinates and on
the light-front}
\label{tab:can}
\end{table}
%\newpage

In the next step one constructs the Hamilton operator and the other
components of the momentum vector. Thus one has completely specified the
dynamics and can start solving the equations of motion.
Typically, one either makes some variational ansatz or a Fock space expansion.
In the latter approach one writes the hadron wave function
as a sum over components with a fixed number of elementary quanta
(for example in QCD: $q\bar{q}$, $q\bar{q}q\bar{q}$, $q\bar{q}g$, e.t.c.).
The expansion coefficients, i.e. the wavefunctions for the corresponding
Fock space sector are used as variational parameters. They are determined
by making the expectation value of the energy stationary
with respect to variations in the wavefunction. Typically the variation is done
for fixed momentum.\footnote{On 
the LF this is very important because
$P_+ \propto 1/P_-$, i.e. unrestricted variation
($P_-$ allowed to vary) results in $P_-\rightarrow \infty$.}
This whole procedure results in coupled integral equations
for the Fock space components. In general they have to be solved
numerically. In practical calculations, since one cannot
include infinitely many Fock components, one has to introduce
some {\it ad hoc} cutoff in the Fock space. Thus it is very important to
demonstrate that physical observables do not depend on how many
Fock components are included.

Until one selects the canonically conjugate momenta and postulates equal $\tau$ commutation relations, 
i.e. at the level of the classical Lagrangian, 
the transition from ET to the LF consists of a mere rewriting. 
After quantization, the independent degrees
of freedom are the fields and their conjugate
momenta on the initial surface ($x^0=0$ for ET and
$x^+=0$ for LF). Thus different degrees of freedom are
employed to expand physical states in the ET
and in the LF approach. Of course, after solving the equations of motion,
physical observables must not depend on the choice of quantization plane.
However, it may turn out that one approach is more efficient
(e.g. faster numerical convergence) than the other or more elegant and
more easy to interpret physically. This is particularly the case if one
is mostly interested in observables which are dominated by correlation
functions in a light-like direction, such as parton distributions, which
are much more easily accessible on the LF than in usual coordinates. 

\section{A First Look at the Light-Front Vacuum}
In the Fock space expansion one starts from the vacuum as the ground
state and constructs physical hadrons by successive application 
of creation operators.
In an interacting theory the vacuum is in general an extremely
complicated state and not known a priori. Thus, in general,
a Fock space expansion is not practical because one does not
know the physical vacuum (i.e. the ground state of the
Hamiltonian). In normal coordinates, particularly
in the Hamiltonian formulation, this is a serious obstacle
for numerical calculations.
As is illustrated in Table \ref{tab:vac}, the LF formulation
provides a dramatic simplification at this point.
\begin{table}
%\vspace{5.cm}
\begin{center}
\begin{tabular*}{11.5cm}[t]{@{\extracolsep{\fill}}c|c}
normal coordinates & light-front \\
\rule{6cm}{0.cm}&\rule{6cm}{0.cm}\\
\hline
\multicolumn{2}{c}{free theory}\\[1.5ex]
%\vspace{5.cm}
\multicolumn{2}{c}{
\setlength{\unitlength}{0.6mm}
\special{em:linewidth 0.4pt}
\def\empoint##1{\special{em:point ##1}}
\def\emline##1##2{\special{em:line ##1,##2}}
\begin{picture}(160,65)
\put( 0, 0){\begin{picture}(80,60)( 0, 0)
\put(5,0){\line(1,0){50}}
\put(30,-3){\line(0,1){50}}
\put(60,0){\makebox(0,0){$P_z$}}
\put(75,0){\line(0,1){60}}
\put(30,55){\makebox(0,0){$P^0 = \sqrt{m^2+{\vec{P}}^2}$}}
\put( 5, 27){\circle*{.1}}
\put( 5,27){\circle*{.1}}
  \put(10,22){\circle*{.1}}
  \put(15,18){\circle*{.1}}
  \put(20,14){\circle*{.1}}
  \put(25,11){\circle*{.1}}
  \put(30,10){\circle*{.1}}
  \put(35,11){\circle*{.1}}
  \put(40,14){\circle*{.1}}
  \put(45,18){\circle*{.1}}
  \put(50,22){\circle*{.1}}
  \put(55,27){\circle*{.1}}
\end{picture}}
\put(80, 0){\begin{picture}(80,60)
\put(5,0){\line(1,0){50}}
\put(10,-3){\line(0,1){50}}
\put(60,0){\makebox(0,0){$P_-$}}
\put(10,55){\makebox(0,0)[l]{$P_+ = 
\frac{\T m^2+{\vec P}_{\perp}^2}{\T 2 P_-}$}}
\put(15,49){\circle*{.1}}
\put(20,25){\circle*{.1}}
\put(25,17){\circle*{.1}}
\put(30,13){\circle*{.1}}
\put(35,10){\circle*{.1}}
\put(40,8){\circle*{.1}}
\put(45,7){\circle*{.1}}
\put(50,6){\circle*{.1}}
\put(55,6){\circle*{.1}}
\end{picture}}
\end{picture}
}\\[1.cm]
$ P^0 = {\D\sum\limits_{\vec{k}}} a^\dagger_{\vec{k}} a_{\vec{k}} \sqrt{m^2 + \vec{k}^2 } $ &
$ P_+ = {\D\sum\limits_{k_-,{\vec k}_{\perp}}} 
a^\dagger_{k_-\!,{\vec k}_{\perp}} a_{k_-\!,{\vec k}_{\perp}}
 \frac{ m^2 + {\vec k}_{\perp}^2 }{ 2 k_-} $ \\[1.5ex]
\multicolumn{2}{c}{vacuum (free theory)}\\[1.5ex]
$\D a_{\vec{k}}|0\rangle = 0 $ & $\D a_{k_-\!,k_\perp}|0
\rangle = 0 $\\[1.5ex]
\multicolumn{2}{c}{vacuum (interacting theory)}\\[1.5ex]
many states with $\vec{P}=0$ & $k_- \ge 0$ \\
(e.\,g.\ $a_{\vec{k}}^\dagger
 a_{-\vec{k}}^\dagger|0\rangle$) &
$\hookrightarrow$ only pure zero-mode \\
 & excitations have $P_-=0$\\[1.5ex]
$\hookrightarrow$ $|\tilde{0}\!>$ very complex &
$\hookrightarrow$ $|\tilde{0}\!>$ can only contain \\
 & zero-mode excitations
\end{tabular*}
\end{center}
\caption{Zero Modes and the Vacuum}
\label{tab:vac}
\end{table}
While all components of the momentum in normal coordinates can
be positive as well as negative, the longitudinal LF momentum
$P_-$ is always positive. In free field theory (in normal
coordinates as well as on the LF) the vacuum is the state which
is annihilated by all annihilation operators $a_k$.
In general, in an interacting theory, excited states (excited with
respect to the free Hamiltonian) mix with the trivial vacuum 
(i.e. the free field theory vacuum) state
resulting in a complicated physical vacuum.
Of course, there are certain selection rules and only states with
the same quantum numbers as the trivial vacuum can mix with
this state; for example, states with the same momentum as
the free vacuum (${\vec P}=0$ in normal coordinates,
$P_-=0$, ${\vec P}_\perp =0$ on the LF).
In normal coordinates this has no deep consequences because there
are many excited states which have zero momentum. On the LF
the situation is completely different. Except for pure zero-mode
excitations, i.e. states where only the zero-mode
(the mode with $k_-=0$) is excited, all excited states have
positive longitudinal momentum $P_-$. Thus only these pure zero-mode
excitations can mix with the trivial LF vacuum.
Thus with the exception of the zero-modes the physical LF vacuum
(i.e. the ground state) of an interacting
field theory must be trivial (the only exceptions are pathological
cases, where the LF Hamiltonian is unbounded from below).

Of course, this cannot mean that the vacuum is entirely
trivial. Otherwise it seems impossible to describe
many interesting problems which are related to spontaneous
symmetry breaking within the LF formalism. For example one knows
that chiral symmetry is spontaneously broken in QCD
and that this is responsible for the relatively small mass of
the pions --- which play an important role in strong interaction
phenomena at low energies. What it means is that one has
reduced the problem of finding the LF vacuum to the problem
of understanding the dynamics of these zero-modes.

First this sounds just like merely shifting the problem
about the structure of the vacuum from nonzero-modes
to zero-modes. However, as the
free dispersion relation on the LF, 
\begin{equation}
k_+=\frac{m^2+{\vec k}^2_{\perp}}{2k_-},
\end{equation}
indicates, zero-modes are high energy modes!
Hence it should, at least in principle, be possible
to eliminate these zero-modes systematically
giving rise to an effective LF field theory.

\section{Effective Light-Front Hamiltonians}

Above discussion shows that it is clearly wrong to leave out zero-modes 
degrees of freedom in the sense of just dropping them since this always yields
a trivial vacuum. However, this should be distinguished from
leaving out zero-modes in the sense of integrating them out. In the
second approach the vacuum can be nontrivial, but the nontrivial
vacuum structure has been shifted from the states to the operators.
In particular, LF Hamiltonians will in general have to be modified
from their canonical form. That this task can actually be accomplished
will be illustrated in the following examples.

\subsection{The 't Hooft Model}
The first model that we will discuss is 1+1 dimensional QCD
for an infinite number of colors (the 't Hooft model).
The model was first solved by 't Hooft in Ref. \cite{th:qcd}, 
using naive LF formulation, i.e. starting from the canonical
LF Hamiltonian and without zero-modes. Nevertheless, 't Hooft's
spectrum was later confirmed by a calculation which employed
canonical equal time quantization \cite{qcd2:et}.
This result is at first very surprising when one considers the
fact that the equal time quantized result yielded a nonzero
chiral condensate for vanishing quark mass $m_{q}\rightarrow 0$, 
while the LF-vacuum was trivial for any value of $m_q$.
a direct evaluation gave of course zero on the LF.
However, application of current algebra techniques to meson masses 
and coupling constants determined by solving the (zero-mode free)
LF equations gives a nonzero result for the condensate
--- even in the zero quark mass limit:
\begin{eqnarray}
0 &=& \lim_{q \rightarrow 0} iq^\mu \int d^2x e^{iqx}
\langle 0|T\left[\bar{\psi}\gamma_\mu \gamma_5\psi (x)
\bar{\psi}i\gamma_5\psi (0)\right] |0\rangle
\nonumber\\
&=&-\langle 0|\bar{\psi}\psi |0\rangle
-2m_q \int d^2x 
\langle 0|T\left[\bar{\psi}i \gamma_5\psi (x)
\bar{\psi}i\gamma_5\psi (0)\right] |0\rangle .
\nonumber\\
\label{eq:cural}
\end{eqnarray}
Upon inserting a complete set of meson states \footnote{
Because we are working at leading order in $1/N_C$, the
sum over one meson states saturates the operator product
in Eq.(\ref{eq:cural}).}
one thus obtains
\begin{equation}
\langle 0|\bar{\psi}\psi |0\rangle
= -m_q \sum_n \frac{f_P^2(n)}{M_n^2},
\label{eq:naivesum}
\end{equation}
where 
\begin{equation}
f_P(n)\equiv \langle 0|\bar{\psi}i\gamma_5\psi |n\rangle
=\sqrt{\frac{N_C}{\pi}} \frac{m_q}{2}\int_0^1 dx \frac{1}{x(1-x)} \phi_n(x)\ \ 
\end{equation} 
and the wave functions $\phi_n$ and invariant masses $M_n^2$ are
obtained from solving
't Hooft's bound state equation for mesons in 
$QCD_{1+1}$
\begin{equation}
M_n^2 \phi_n(x) = \frac{m_q^2}{x(1-x)}\phi_n(x) + G^2 \int_0^1 dy
\frac{\phi_n(x)-\phi_n(y)}{(x-y)^2}.
\label{eq:thooft}
\end{equation}
The result for $\langle 0|\bar{\psi}\psi |0\rangle$
obtained this way agrees with the ET calculation \cite{qcd2:lfvac}
\footnote{It should be emphasized that the LF calculation preceded
the ET calculation.}.
\begin{figure}
\unitlength1.cm
\begin{picture}(10,6.)(2.9,14.1)
\includegraphics{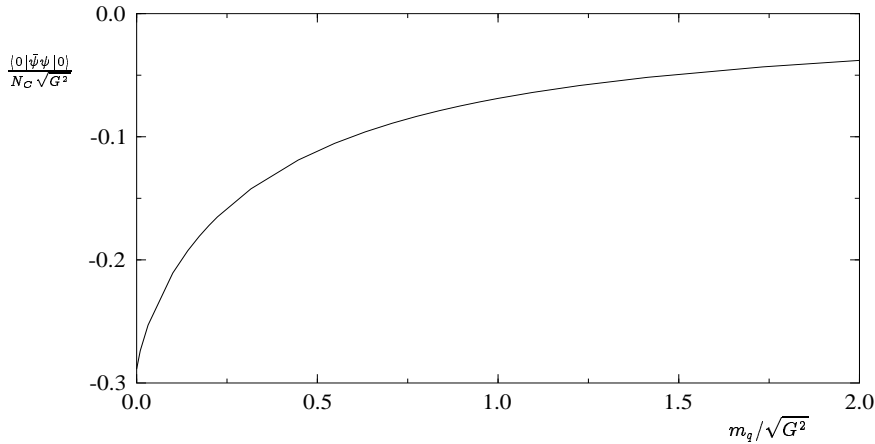}
\end{picture}
\caption{Chiral condensate obtained by evaluating Eq. (\protect\ref{eq:naivesum})
as a function of the quark mass. For nonzero quark mass, the (infinite)
free part has been subtracted. The result agrees for all quark mass with
the calculation done using equal time quantization.
}
\label{fig:thooft}
\end{figure}

This seemingly paradoxical result (peaceful coexistence of a Fock
vacuum and a nonzero fermion condensate) can be understood by 
defining LF quantization through a limiting procedure \cite{le:ap},
where the quantization surface is kept space-like, but being
carefully ``rotated'' to the LF \footnote{Another way of thinking
about this procedure is to imagine a gradual boost to infinite
momentum.}. 
Not all physical quantities behave continuously under this procedure
as the LF is approached. For example, the chiral condensate
$\langle 0|\bar{\psi}\psi|0\rangle $ has a discontinuous LF limit.
On the other hand, the equation of motion for mesons in $QCD_{1+1}$
does have a smooth LF limit. This result explains why the current
algebra relation
gives the right result for the condensate, even though 
$\langle 0|\bar{\psi}\psi|0\rangle $ vanishes when evaluated
directly on the LF: Since the bound state equation for mesons
has a smooth LF limit, both meson masses and coupling constant
can be evaluated directly on the LF. Since the current algebra
relation (\ref{eq:cural}) is a frame independent relation, it can
then be used to extract the condensate from the LF calculation. 
However, since $\langle 0|\bar{\psi}\psi|0\rangle $ has a 
discontinuous LF limit, it would be misleading to draw conclusions
about the vacuum structure from its value obtained directly on the LF.

Unfortunately, the 't Hooft model is rather untypical
because the effective (in the sense of zero-modes integrated out)
LF Hamiltonian for this model is identical to the canonical
Hamiltonian, i.e. the naive calculation yielded already the correct
spectrum. In general, the situation is more complex as we will
illustrate in the following examples.

\subsection{Self-Interacting Scalar Fields}
The most simple example with a non-trivial effective LF Hamiltonian
are self-interacting scalar fields described by a Lagrangian
of the form
\be
{\cal L} = \frac{1}{2}\partial_\mu \phi \partial^\mu \phi
- \sum_{n=2}^N \frac{\lambda_n}{n!} \phi^n .
\ee
We will not restrict the dimensionality of space, but, depending
on the number of space time dimensions, we will limit the
largest power $N$ appearing in the interactions in order to ensure
renormalizability. Furthermore, we will implicitly assume that 
whenever cutoffs are used, compatible cutoffs are used in the
LF and ET frameworks, i.e. for example dimensional regularization
in the $\perp$ direction or a $\perp$ momentum cutoff.

\begin{figure}
\unitlength1.cm
\begin{picture}(10,5.)(-5,0)
\includegraphics{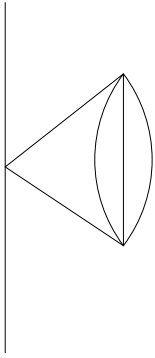}
\end{picture}
\caption{Generalized tadpole (Feynman-) diagram in $\phi^4$ theory.
}
\label{fig:phi4}
\end{figure}

\begin{figure}
\unitlength1.cm
\begin{picture}(10,5.)(-4,-4.5)
\put(-3,-4){\vector(0,1){3.}}
\put(-2.5,-2.5){\makebox(0,0){$x^+$}}
\includegraphics{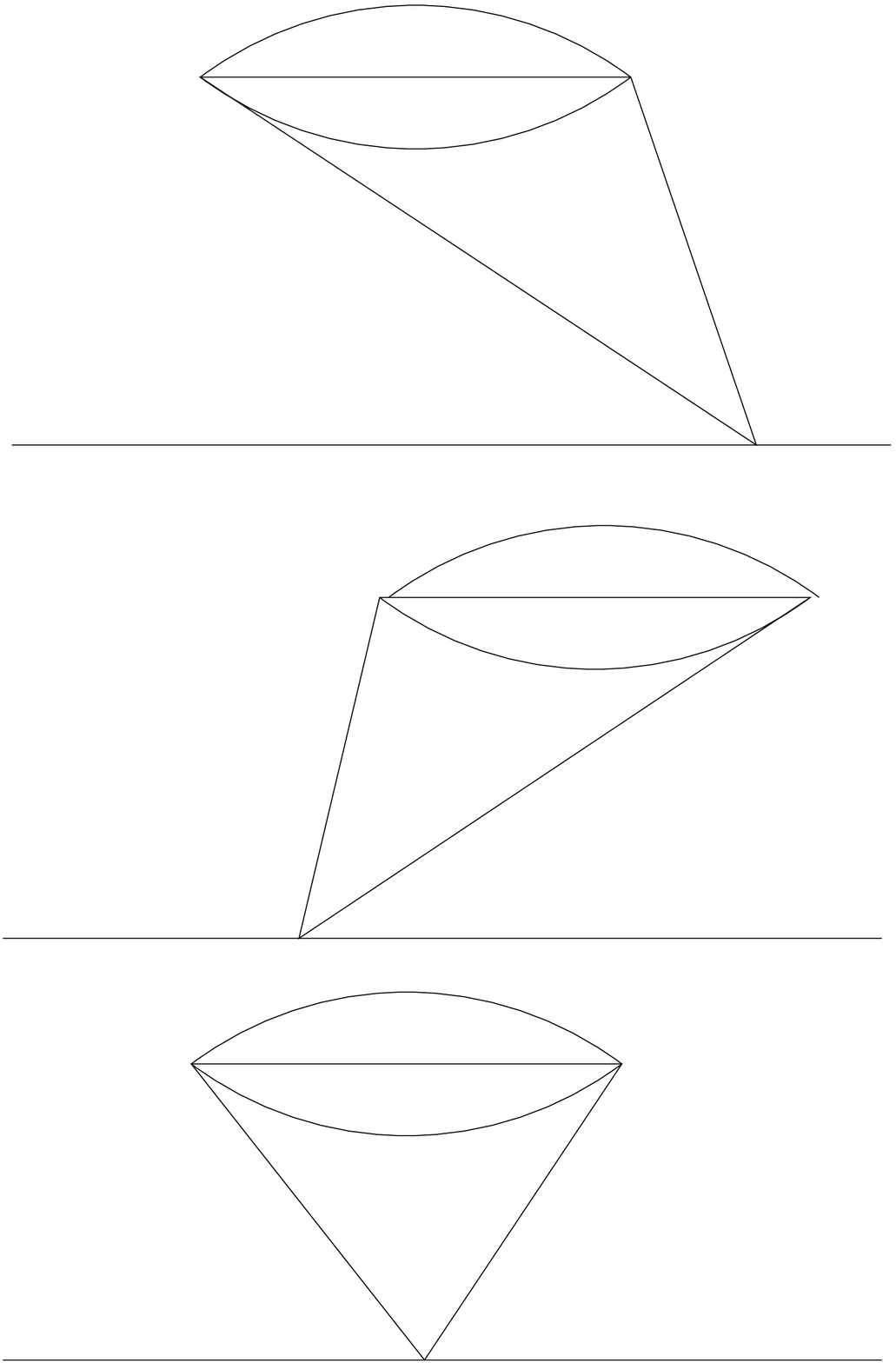}
\end{picture}
\caption{Same as Fig. \protect\ref{fig:phi4} but as LF-time ordered diagrams. At least one of
the vertices has all lines popping out of or disappearing into the vacuum.
}
\label{fig:phi4t}
\end{figure}

The main difference between the LF formulation and the ET formulation
is that generalized tadpoles (a typical example is shown in Fig.\ref{fig:phi4}), 
i.e. Feynman diagrams where one piece
of the diagram is connected to the rest of the diagram only at one
point, cannot be generated by a LF Hamiltonian: in time ordered
perturbation theory, at least one of the
vertices in a generalized tadpole diagram has all lines coming out
of or disappearing into the vacuum (Fig.\ref{fig:phi4t}) --- 
which is forbidden on the LF
(without zero-modes). 

So the bad news is that all generalized tadpole diagrams are zero
on the LF and they are nonzero in ET quantization, i.e. there is
a difference between the perturbation series generated by the two 
formulations \cite{paul:sg,mb:sg}.

The good news is that in self-interacting scalar theories, 
it is only in generalized tadpole
diagrams where such a difference occurs. \footnote{To my knowledge,
there is no strict proof of this result, but it is based on
handwaving arguments as well as on a thorough three loop analysis.} 
Since generalized tadpoles are only constants, their absence can
be compensated for by local counter-terms in the interaction.
Note that generalized tadpoles renormalize only $\phi^n$-couplings
with $n<N$, where $N$ is the largest power appearing in the 
interaction.

From the purely practical point of view, the difference between LF quantization and ET
quantization arises only if one compares calculations done with
the same bare parameters! Suppose one would start with parameters
that have been chosen so that the bare parameters on the LF
include already the counter-terms that are necessary to compensate
for the absence of tadpoles. Then ET and LF formulation should
give the same results for all n-point Green's functions, i.e.
physical observables should be the same. But how can one find
the appropriate counter-terms without having to refer to an ET
calculation? There is nothing easier than that: simply by using
only physical input parameters to fix the bare parameters!
For example, if one matches the physical masses of a few particles
between an ET calculation and a LF calculation then the bare parameters
that one needs in order to get these masses will be different in ET
and LF quantization. The difference will be just such that it compensates
for the absence of tadpoles on the LF and hence all further 
observables will be the same. In other words there is no problem
at all with the LF formulation --- despite the absence of all
generalized tadpole diagrams. The only situation where one needs to
be a little careful in the LF approach is a situation where
spontaneous symmetry breaking occurs in the ET formulation.
In this case, one has to allow for odd terms in $\phi$ in the
effective LF Hamiltonian, while the ET Hamiltonian contains only
even terms.

Beyond this happy ending, there is one more very interesting aspect
to this story, which has to do with vacuum condensates. For example,
every generalized tadpole diagram in $\phi^4$ theory is numerically
equal to a diagram that contributes to $\langle 0|\phi^2|0\rangle$.
For example, the tadpole in Fig. \ref{fig:phi4} is proportional to
a term that contributes to $\langle 0|\phi^2|0\rangle$ to second order in $\lambda$.
In fact, after working out the details one finds that the
additional LF counter-term, necessary to obtain equivalence between
ET and LF quantization is a mass counter-term \cite{mb:sg}
\begin{equation}
\Delta m^2 =\lambda \langle 0|\frac{:\phi^2:}{2}|0\rangle ,
\end{equation}
where $\lambda$ is the four point coupling and the vacuum expectation
value (VEV) on the r.h.s. is to be evaluated in normal coordinates.

This result can be readily generalized to an arbitrary polynomial
interaction. One finds the following dictionary: perturbation theory
based on a canonical equal time Hamiltonian with
\begin{equation}
{\cal L}_{int}^{ET}=\sum_n \lambda_n \frac{:\phi^n:}{n!} 
\end{equation} 
and perturbation theory based on a canonical light-front
Hamiltonian with
\begin{equation}
{\cal L}_{int}^{LF}=\sum_n \tilde{\lambda}_n \frac{:\phi^n:}{n!} 
\end{equation} 
are equivalent if 
\begin{equation}
\tilde{\lambda}_n = \sum_{k\leq n} \lambda_{n-k} 
\langle 0|\frac{:\phi^k:}{k!}|0\rangle  .
\label{eq:dict}
\end{equation}
\begin{figure}
\unitlength1.cm
\begin{picture}(10,4.)(1,-7.5)
\includegraphics{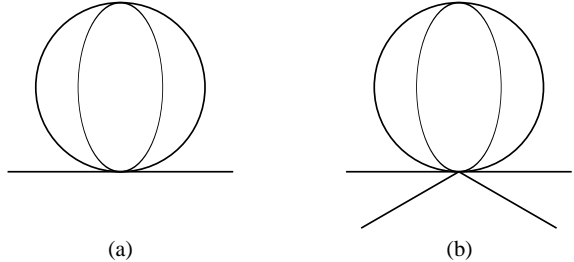}
\end{picture}
\caption{Generalized tadpole diagrams for scalar field theories with
higher polynomial interactions. Both are set to zero in LF quantization without
zero-modes. Both are proportional to $\langle 0|\phi^4|0\rangle$.
The diagram in a.) gives rise to a mass renormalization counter-term and b.)
renormalizes the four-point interaction.
}
\label{fig:basket}
\end{figure}
In Ref.\cite{mb:sg} this fundamental result was derived perturbatively
and the prescription for constructing the effective Hamiltonian was
tested non-perturbatively by calculating physical masses of
of excited ``mesons'' and solitons in the sine-Gordon model.

It is very tempting to conjecture that this dictionary
(\ref{eq:dict}) also holds for
non-perturbative condensates (such as condensates which arise after
spontaneous symmetry breaking). While a general proof is still
missing, it has indeed been possible to demonstrate for a few
specific models that the conjecture is correct \cite{naus:sg}.

It should also be emphasized that these equivalence considerations
hold irrespective of the number of space-time dimensions, i.e. they
apply to 1+1 as well as 2+1 and 3+1 dimensional theories. One
must only be careful to use commensurate cutoffs when comparing
ET and LF quantized theories. An example would be a transverse 
lattice cutoff, which can be employed both in ET quantization as well
as in LF quantization.

What makes all these results particularly interesting is that 
the effective (zero-modes integrated out) couplings in
the effective LF-Hamiltonians contain condensates which shows
how non-perturbative effects can
``sneak'' into the LF formalism and how one can resolve the apparent
conflict between trivial LF vacua and nontrivial vacuum effects.

\subsection{Fermions with Yukawa Interactions}
\begin{figure}
\unitlength1.cm
\begin{picture}(10,4.)(1,5)
\includegraphics{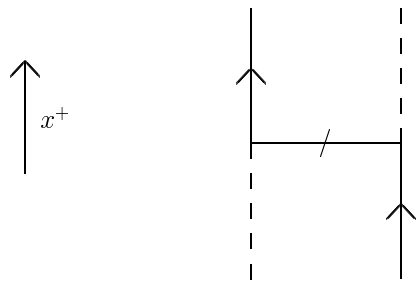}
\end{picture}
\caption{Four point interaction in Yukawa theory that arises after
eliminating $\psi_{(-)}$. The crossed out full line represents the
instantaneous fermion exchange interaction that results from this
elimination. Contracting the boson lines (dashed) yields
a diagram analogous to the tadpole diagrams for self-interacting scalar
fields. 
}
\label{fig:inst}
\end{figure}

Eventually, we are interested to understand chiral symmetry breaking in QCD, 
i.e. we need to understand fermions. As a first step in this direction
let us consider a Yukawa model in 1+1 dimensions
\begin{equation}
{\cal L}=\bar{\psi}\left( i\not \!\!\partial 
-m_F-g\phi \gamma_5 \right)\psi -\frac{1}{2}
\phi\left( \Box +m_B^2\right)\phi .
\end{equation}
The main difference between scalar and Dirac fields in the LF formulation is
that not all components of the Dirac field are dynamical: multiplying the
Dirac equation
\begin{equation}
\left( i\not \!\!\partial 
-m_F-g\phi \gamma_5\right)\psi =0
\end{equation}
by $\frac{1}{2}\gamma^+$ yields a constraint equation (i.e. an
``equation of motion'' without a time derivative)
\begin{equation}
i\partial_-\psi_{(-)}=\left(m_F+g\phi\gamma_5\right)\gamma^+\psi_{(+)}
,
\label{eq:constr}
\end{equation}
where
\begin{equation}
\psi_{\pm}\equiv \frac{1}{2}\gamma^\mp \gamma^\pm \psi .
\end{equation}
For the quantization procedure, it is convenient to eliminate
$\psi_{(-)}$, using
\be
\psi_{(-)} = \frac{\gamma^+}{2i\partial_-}\left(m_F+g\phi\gamma_5\right) 
\psi_{(+)}  
\ee
from the classical Lagrangian before imposing
quantization conditions, yielding
\begin{eqnarray}
{\cal L}&=&\sqrt{2}\psi_{(+)}^\dagger \partial_+ \psi_{(+)}
-\phi\left( \Box +m_B^2\right)\phi
-\psi^\dagger_{(+)}\frac{m_F^2}{\sqrt{2}i\partial_-}
\psi_{(+)}
\label{eq:lelim}
\\
&-&\psi^\dagger_{(+)}\left(
g\phi
\frac{m_F\gamma_5}{\sqrt{2}i\partial_-}
+\frac{m_F\gamma_5}{\sqrt{2}i\partial_-}g\phi\right)
\psi_{(+)}
-\psi^\dagger_{(+)}g\phi\frac{1}{\sqrt{2}i\partial_-}
g\phi\psi_{(+)} .
\nonumber
\end{eqnarray}
The rest of the quantization procedure very much resembles the procedure
for self-interacting scalar fields.
In particular, one must be careful about generalized tadpoles,
which might cause additional counter-terms in the LF Hamiltonian.
In the Yukawa model one usually (i.e. in a covariant formulation)
does not think about tadpoles. However, after eliminating $\psi_{(-)}$,
we are left with a four-point interaction in the Lagrangian, which does
give rise to time-ordered diagrams that resemble tadpole diagrams 
(Fig.\ref{fig:inst}).
In fact, the four-point interaction gives rise to diagrams where
a fermion emits a boson, which may or may not self-interact, and then
re-absorb the boson at the same LF-time. \footnote{There are also tadpoles,
where the fermions get contracted. But those only give rise to an additional
boson mass counter-term, but not to a non-covariant counter-term that we
investigate here.}
As we discussed already in detail in the context of self-interacting
scalar fields, such interactions
cannot be generated by a LF Hamiltonian, i.e. the LF formalism
defines such tadpoles to be zero.

For self-interacting scalar fields, the difference between ET and LF
perturbation theory which thus results can be compensated by a
redefinition of parameters that appear already in the Lagrangian.
In the Yukawa model, the situation is a little more complicated.
The missing tadpoles have the same operator/Lorentz structure as the
so called kinetic mass term
\begin{equation}
{\cal P}^-_{kin} =\psi^\dagger_{(+)}\frac{m_F^2}{\sqrt{2}i\partial_-}
\psi_{(+)} .
\label{eq:pkin}
\end{equation}
One obtains this result by contracting the two scalar fields in the
four-point interaction. More details can be found in Ref. \cite{mb:alex}.
The important point here is that there is no similar counter-term
for the term linear in the fermion mass $m_F$. Thus the difference
between ET and LF quantization cannot be compensated by tuning the
bare masses differently. The correct procedure requires to renormalize
the kinetic mass term (the term $\propto m_F^2$) and the vertex mass term
(the term $\propto m$) independent from each other. More explicitly this
means that one should make an ansatz for the renormalized LF Hamiltonian
density of the form
\begin{eqnarray}
{\cal P}^-&=&
\frac{m_B^2}{2}\phi^2
+\psi^\dagger_{(+)}\frac{c_2}{\sqrt{2}i\partial_-}
\psi_{(+)}
+c_3\psi^\dagger_{(+)}\left(
\phi
\frac{\gamma_5}{\sqrt{2}i\partial_-}
+\frac{\gamma_5}{\sqrt{2}i\partial_-}g\phi\right)
\psi_{(+)} \nonumber\\
&+&c_4\psi^\dagger_{(+)}\phi\frac{1}{\sqrt{2}i\partial_-}
\phi\psi_{(+)} ,
\label{eq:pren}
\end{eqnarray}
where the $c_i$ do not necessarily satisfy the canonical relation
$c_3^2=c_2c_4$. However, this does not mean that the $c_i$ are completely
independent from each other. In fact, only for specific combinations of
$c_i$ will Eq.(\ref{eq:pren}) describe the Yukawa model. It is only that
we do not know the relation between the $c_i$.\footnote{It is conceivable that
coupling constant coherence \cite{brazil} might also be helpful in finding
the ``unknown relation'', but I have not been able to figure out how to use
this idea in the context of a superrenormalizable model such as above model.}

Thus the bad new is that the number of parameters in the LF Hamiltonian
has increased by one (compared to the Lagrangian). The good news is that
a wrong combination of $c_i$ will in general give rise to a parity 
violating theory. \footnote{As an example, consider the free theory,
where the correct relation ($c_3^2=c_2c_4$) follows from a covariant
Lagrangian. Any deviation from this relation can be described on
the level of the Lagrangian (for free massive fields, equivalence
between LF and covariant formulation is not an issue)
by addition of a term of the form $\delta {\cal L} =
\bar{\psi}\frac{\gamma^+}{i\partial^+}\psi$, which is obviously 
parity violating, since parity transformations result in
$A^\pm \rightarrow A^\mp$ for Lorentz vectors $A^\mu$; i.e. $\delta {\cal L}
\rightarrow \bar{\psi}\frac{\gamma^-}{i\partial^-}\psi \neq \delta {\cal L}$.}
This is good news because one can thus use parity
invariance for physical observables as an additional renormalization
condition to determine the additional ``free'' parameter.

In fact, in Ref. \cite{mb:parity}, it was shown that 
utilizing parity constraints as non-perturbative 
renormalization conditions
is practical. The observable considered in that work was the
vector transition form factor (in a scalar Yukawa theory in 1+1 dimensions) 
between physical meson states of opposite C-parity (and thus
supposedly opposite parity)

\begin{equation}
\langle p^\prime, n| j^\mu | p, m \rangle
\stackrel{!}{=} \varepsilon^{\mu \nu}q_\nu F_{mn}(q^2),
\label{eq:form}
\end{equation}
where $q=p^\prime -p$. When writing the r.h.s. in terms of
one invariant form factor, use was made of both vector current
conservation and parity invariance. A term proportional
to $p^\mu + {p^\prime}^\mu$ would also satisfy current conservation,
but has the wrong parity. A term proportional to 
$\varepsilon^{\mu \nu}\left(p_\nu + p_\nu^\prime\right)$ 
has the right parity,
but is not conserved and a term proportional to $q^\mu$ is both
not conserved and violates parity. Other vectors do not exist for
this example.
The Lorentz structure in Eq. (\ref{eq:form}) has nontrivial
implications even if we consider only the ``good'' component
of the vector current \footnote{In the context of LF calculations,
currents that are bilinear in the dynamical component $\psi_{(+)}$
are usually easiest to renormalize and calculate. Other combinations,
such as $\psi_{(-)}^\dagger \psi_{(-)}$ involve interactions when
expressed in terms of the dynamical components and are thus terrible
to handle --- hence they are often referred to as ``bad'' components.},
yielding
\begin{equation}
\frac{1}{q^+}\langle p^\prime, n| j^+ | p, m \rangle
=F_{mn}(q^2).
\label{eq:formplus}
\end{equation}
That this equation implies nontrivial constraints can be seen
as follows: as a function of the longitudinal momentum transfer
fraction $x\equiv q^+/p^+$, the invariant momentum transfer 
reads ($M_m^2$ and $M_n^2$ are the invariant masses of the
in and outgoing meson)
\begin{equation}
q^2=x\left(M_m^2 -\frac{M_n^2}{1-x}\right)
\end{equation}
\begin{figure}
\unitlength1.cm
\begin{picture}(10,12)(0,1.)
\includegraphics{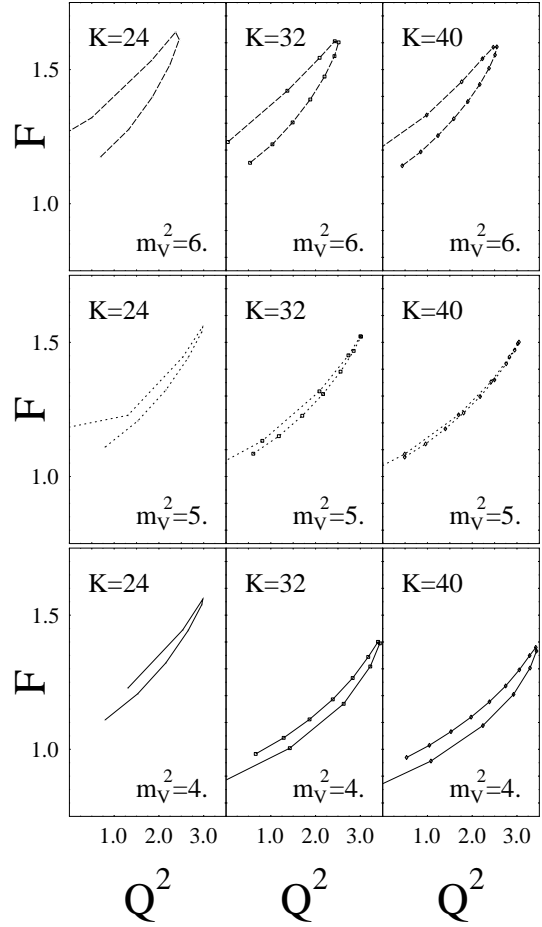}
\end{picture}
\caption{
Inelastic transition form factor (\protect\ref{eq:formplus}) between the
two lightest meson states of the Yukawa model, calculated
for various vertex masses $m_v$ and for various DLCQ parameters
$K$. The physical masses for the fermion and the scalar meson
have been renormalized to the values 
$\left(m_F^{phys}\right)^2=\left(m_F^{phys}\right)^2=4$.
All masses and momenta are in units of
$\protect\sqrt{\lambda} = \protect\sqrt{c_4/2\pi}$. In this example,
only for $m_V^2\approx 5$ one obtains a form factor that is a unique
function of $Q^2$, i.e. only for $m_V^2\approx 5$, the result is
consistent with Eq. (\protect\ref{eq:formplus}). Therefore, only for
this particular value of the vertex mass, is the matrix element of the current operator consistent with both parity and current conservation.
}
\label{fig:parity}
\end{figure}
Typically, there are two values of $x$ that lead to the same
value of $q^2$. It is highly nontrivial to obtain the same
form factor for both values of $x$. In Ref. \cite{mb:parity},
the coupling as well as the 
physical masses of both the fermion and the lightest boson where
kept fixed, while the ``vertex mass'' was tuned (note that this
required re-adjusting the bare kinetic masses). Figure \ref{fig:parity}
shows a typical example. In that example, the calculation of the
form factor was repeated for three values of the DLCQ parameter K
(24, 32 and 40) in order to make sure that numerical approximations
did not introduce parity violating artifacts.

For the ``magic value''
of the vertex mass one finds that the parity condition 
(\ref{eq:formplus}) is satisfied over the whole range of $q^2$
considered. This provides a strong self-consistency check, since
there is only one free parameter, but the parity condition is not
just one condition but a condition for every single value of
$q^2$ (i.e. an infinite number of conditions). In other words,
keeping the vertex mass independent from the kinetic mass is not
only necessary, but also seems sufficient in order to properly 
renormalize Yukawa$_{1+1}$.

However, note that, depending on the cutoff used, the kinetic mass 
counter-term may or may not be a function of the momenta \cite{mb:finite}.
If one uses a cutoff which breaks manifest boost invariance,
such as a DLCQ cutoff, then the kinetic mass counter-term will in
general have to be a function of the longitudinal momenta.
On the other hand, if one uses cutoffs which preserve manifest boost
invariance, such as cutoffs on ratios of momenta or invariant mass
transfers at each vertex then the kinetic mass counter-term can be
taken to be momentum independent and thus counter-term functions
can be avoided in this model. 

It is still an open question whether
``counter-term functions'' \cite{osu} can be avoided in general, i.e. 
in 3+1 dimensions. In non-gauge theories, preliminary investigations
(see also Section \ref{sec:3dmodel}
suggest that counter-term functions can be avoided
if one uses a cutoff where longitudinal and transverse momenta are
cut off independently and the cutoff must at the same time be manifestly
boost invariant in the longitudinal direction. For gauge theories,
the same condition seems to hold, except that the transverse momenta must
be cut off in a gauge invariant way (to avoid disastrous longitudinal
divergences) by using dimensional regularization or a transverse lattice.

\subsection{A 3+1 Dimensional Model with Spontaneous Chiral
Symmetry Breaking}
\label{sec:3dmodel}
One would like to study a 3+1 dimensional model which goes
beyond the mean field approximation (NJL !), but on the other hand
being too ambitious results in very difficult or unsolvable models.
\footnote{For example, demanding Lorentz invariance, chiral symmetry and 
asymptotic freedom leaves QCD as the most simple model.}
We decided to place the following 
constraints on our model:\cite{mb:hala} 
\begin{itemize}
\item Most importantly, the model should be 3+1 dimensional,
but we do not require full rotational invariance.
\item The model should have spontaneous $\chi$SB  (but not just mean field)
\item Finally, it should be solvable both on the LF and using a conventional
technique (to provide a reference calculation).
\end{itemize}

Given these constraints, the
most simple model that we found is described by the Lagrangian
\begin{eqnarray}
{\cal L} = \bar{\psi_k}\left[ \delta^{kl}\left(i\partial\!\!\!\!\!\!\not \;\; -
m  \right)
-\frac{g}{\sqrt{N_c}}{\vec \gamma}_\perp {\vec A}^{kl}_\perp \right]\psi_l -
\frac{1}{2}
{\vec A}^{kl}_\perp \left(\Box +\lambda^2\right){\vec A}^{kl}_\perp ,
\label{eq:lager}
\end{eqnarray}
where $k,l$ are ``color'' indices ($N_c\rightarrow \infty$),
$\perp =x,y$ and where a
cutoff is imposed on the transverse momenta. A fermion mass was introduced
to avoid pathologies associated with the strict $m=0$ case.
$\chi SB$ can be studied by considering the $m\rightarrow 0$ limit of the
model.

The reasons for this bizarre choice of model [Eq. (\ref{eq:lager})] 
are as follows.
If one wants to study spontaneous breaking of chiral symmetry, then
one needs to have a chirally invariant interaction to start with, which
motivates a vector coupling between fermions and bosons. However, we
restricted the vector coupling to the $\perp$ component of a vector field
since otherwise one has to deal with couplings to the ``bad'' current
$j^-$ \footnote{$j^-$ is bilinear in the constrained component of the
fermion field, which makes it very difficult to renormalize this component
of the current in the LF framework.}. 
In a gauge theory, such couplings can be avoided by choice of gauge, but
we preferred not to work with a gauge theory, since this would give rise
to additional complications from infrared divergences.
Furthermore, we used a model with ``color'' degrees of freedom and
considered the limit where the number of colors is infinite, because such
a model is solvable, both on and off the LF. No interaction among the bosons
was included because this would complicate the model too much.
Finally, we used a cutoff on the transverse momenta because such a cutoff can
be used both on the LF as well as in normal coordinates and therefore
one can compare results from these two frameworks already for finite
values of the cutoff.  

Because we are considering the limit 
$N_C \rightarrow \infty$, of Eq. (\ref{eq:lager}), the iterated
rainbow approximation 
for the fermion self-energy $\Sigma$ becomes exact, which we used
to solve the model in the Dyson-Schwinger approach. On the one hand this 
provides a reference calculation, but one can also use the solution
to show that for sufficiently large coupling constant,
the physical mass for the fermion remains finite in the limit
$m \rightarrow 0$, proving the spontaneous breakdown of chiral symmetry
in the model.

Since we wanted to investigate the applicability of the effective LF 
Hamiltonian formalism, we formulated above model on the LF without explicit
zero-mode degrees of freedom. The bottom line of this investigation
was as follows (for details see Ref. \cite{mb:hala}).

\begin{itemize}
\item The LF Green's function equation and the DS equation are identical 
(and thus have identical solutions) if and only if
one introduces an additional (in addition to the self-induced inertias)
counter-term to the kinetic mass term for the fermion.
\item For fixed transverse momentum cutoff, this additional kinetic
mass term is finite.
\item The value of the vertex mass in the LF Hamiltonian is the same as 
the value of the current mass in the DS equation.
\item In the chiral limit, mass generation for the (physical)
fermion occurs through
the kinetic mass counter term
\item The effective interaction
(after integrating out 0-modes) can be summarized by a
few simple terms --- which are already present in the canonical 
Hamiltonian.
\end{itemize}

Even though we determined the kinetic mass counter term by directly
comparing the LF and DS calculation, several methods are conceivable which 
avoid reference to a non-LF calculation in order to set up the LF problem.
One possible procedure would be to impose parity invariance for physical
observables as a constraint \cite{mb:parity}.

\section{Dynamical Vertex Mass Generation and Chiral Symmetry Breaking
on the LF}
Naively, helicity flip amplitudes for fermions
seem to vanish in the chiral limit of
light-front QCD, which would make it nearly impossible to generate a
small pion mass in this framework. In this Section, a simple model is
used to illustrate
how a large helicity flip amplitude is generated dynamically by summing
over an infinite number of Fock space components. Implications for the
renormalization of light-front Hamiltonians for fermions are discussed.

In the LF-formulation, only half of the spinor components are dynamical
degrees of freedom in the sense that their equation of motion 
involves a time derivative. Upon introducing 
$\psi_{(\pm)}=\gamma^\pm \gamma^\mp \psi/2$, one finds for example in
QCD that $\psi_{(-)}$ satisfies a constraint equation \cite{osu}
\begin{equation}
i\partial_- \psi_{(-)} = \left[ {\vec \alpha}_\perp \cdot
\left( i{\vec \partial}_\perp + g {\vec A}_\perp\right) + \gamma^0m_F
\right] \psi_{(+)}
\end{equation}
and $\psi_{(-)}$ is usually eliminated (using this constraint equation)
from the Lagrangian before quantizing the theory.
Thus the Hamiltonian contains both a term quadratic in the fermion mass
(the kinetic energy term for the fermions) and one term which is linear
in the fermion mass (one gluon vertex with helicity flip).

It has been known for a long time that integrating out zero-mode
degrees of freedom results in a nontrivial renormalization of the
quadratic (kinetic!) mass term \cite{mb:rot} but the linear (helicity flip 
vertex!) mass term in the Hamiltonian is unaffected by strict zero modes
and the ``vertex mass'' must be identified with the current quark mass
which vanishes in the chiral limit (see the previous section as well as
Refs: \cite{osu,mb:hala,ken}).

It is very easy to see how a constituent quark picture can emerge in such
an approach.
However, it always seemed mysterious how one can obtain a massless $\pi$
meson in such a picture without having at the same time a massless $\rho$:
when the helicity flip term for quarks is omitted, $\pi$ and $\rho$ become
partners in a degenerate multiplet \cite{ken}. The key observation to resolve
this problem is that one needs to find a mechanism which dynamically 
generates a large helicity flip amplitude. At first this seems impossible
since every LF-time ordered diagram (any order in the coupling!) which flips 
the helicity of the fermion contains at least one power of 
the vertex mass (which vanishes in the chiral limit).

Above argument does not rule out the possibility that summing over
an infinite class of time-ordered diagrams (including an infinite
number of Fock space components) can lead to divergences
which can compensate for the suppression of individual diagrams.
A simple  model to illustrate this idea is a fermion field with
(fundamental) ``color'' degrees of freedom coupled to the transverse 
components of a massive vector field (adjoint representation) 
\begin{equation}
{\cal L} = \bar{\psi}\left( i \partial\!\!\!\!\!\!\not \;\; - m -
\frac{g}{\sqrt{N_C}}{\vec \gamma}_\perp {\vec A}_\perp \right)\psi - \frac{1}{2}
\mbox{tr} \!\!\left( {\vec A}_\perp \Box {\vec A}_\perp + 
\lambda^2 {\vec A}_\perp^2\right).
\label{eq:model}
\end{equation}
We will take the limit $N_C\rightarrow \infty$, where
the planar approximation becomes exact.
Furthermore, we will assume that the fields depend
on longitudinal coordinates only, i.e. we consider a dimensionally reduced
version of the model. Because of the $N_C\rightarrow \infty$
limit, the model still has spontaneous breaking of chiral symmetry as
$m\rightarrow 0$. This can be shown by solving the covariant self-energy
equation
\begin{equation}
\Sigma (p) = g^2 \int \frac{d^2 k}{(2\pi^2)}
{\vec \gamma}_\perp
\frac{1}{k \!\!\!\!\!\!\not \;\; - m - \Sigma (k)}
{\vec \gamma}_\perp
\frac{1}{(p-k)^2-\lambda^2}
\end{equation}
self-consistently. Note that $\Sigma$ becomes effectively momentum
independent for $\lambda \rightarrow \infty$, yielding a simplified
gap equation
\begin{equation}
M = m + M\frac{g^2M}{2\pi \lambda^2} \ln \frac{\lambda^2}{M^2},
\label{eq:gap1}
\end{equation}
where $M$ is the physical mass of the fermion.

It is easy to see how {\it dynamical vertex mass generation} 
emerges in the 
limit where $\lambda$, the mass of the boson, is very large (in
order to avoid a trivial theory, the coupling constant is rescaled
accordingly). In this limit, self-energies are momentum independent
and one can thus absorb all self-energies
into a re-definition of the kinetic mass, by setting it equal to the 
physical mass $M$, which satisfies the above gap-equation (\ref{eq:gap1}),
i.e. 
\begin{equation}
M = \frac{m}{1-\frac{g^2}{2\pi \lambda^2}\ln \frac{\lambda^2}{M^2}}.
\label{eq:gap}
\end{equation}

However, this does not yet explain
how a large helicity flip amplitude emerges. For this purpose, let us
consider a helicity flip process and focus on all those diagrams which
are linear in the vertex mass, i.e. contain only one helicity flip vertex.
After redefining the kinetic mass as the physical mass, we should no
longer take diagrams with self-mass sub-diagrams 
\footnote{I.e. sub-diagrams 
with only two external fermion lines, which are both on mass shell.}
into account since this would amount to double counting. Without self-mass
sub-diagrams and in the planar approximation considered here, this leaves 
only a very simple class of diagrams which are linear in the vertex mass.
In the following, we will study this class in detail and sum it up to all
orders in perturbation theory.

\begin{figure}
\unitlength1.cm
\begin{picture}(15,6)(-1,-7.5)
\includegraphics{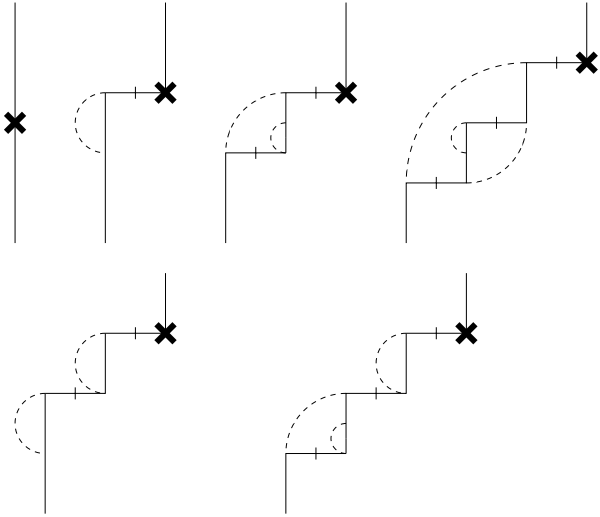}
\end{picture}
\caption{LF-time ordered diagrams contributing to the helicity flip 
amplitude for a fermion. The external vertex is depicted as a cross.
The dashed lines are boson fields and the slashed 
lines represent LF-instantaneous interactions. In the limit of a heavy
boson mass, the diagrams in the top row dominate.
}
\label{fig1}
\end{figure}

\begin{figure}
\unitlength1.cm
\begin{picture}(15,6.5)(-1,-8.5)
\includegraphics{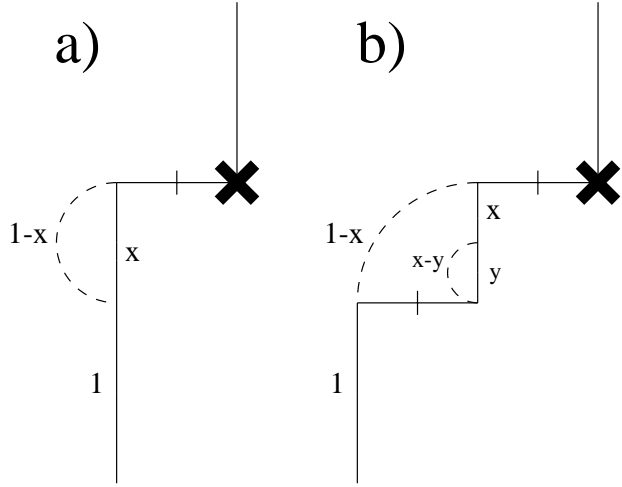}
\end{picture}
\caption{Lowest order bare rainbow diagrams.}
\label{fig:bare}
\end{figure}

\begin{figure}
\unitlength1.cm
\begin{picture}(15,3.2)(.8,-4.5)
\includegraphics{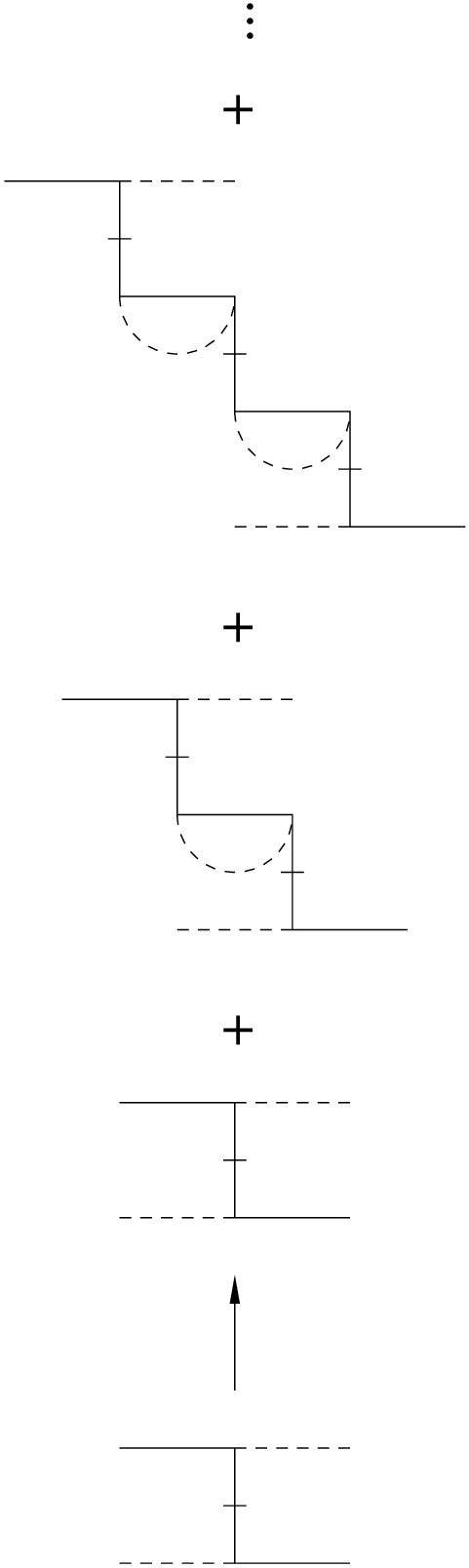}
\end{picture}
\caption{Chain of instantaneous interactions leading to a geometric
series.
}
\label{fig:chain}
\end{figure}

First of all, since we work in a dimensionally reduced model,
boson-fermion vertices are linear in the vertex mass, unless they
are instantaneous vertices, i.e. the limitation to terms linear in the vertex
mass means that all but one vertex in a given time ordered diagram 
must be instantaneous. This means that, to leading order in the vertex mass, 
a general higher order diagram
for a helicity flip vertex is obtained by attaching an instantaneous
interaction to the external vertex and then another one to the boson line
emanating from the first instantaneous vertex and so on until the last
boson line ends up in the actual helicity flip vertex.
Typical diagrams are depicted in Fig. \ref{fig1} 
\footnote{Of course, in addition to
the diagrams in Fig. \ref{fig1}, there are corrections to the helicity
flip amplitude on the other side of the external vertex, but up to
kinematical factors, the are identical to the ones depicted in
Fig. \ref{fig1}.}
Let us first look at the ``bare rainbow'' diagrams \footnote{We call these
diagrams bare rainbow diagrams, since they really have the topology of
a rainbow, as distinguished from iterated or nested rainbows.}
in the top row of Fig. \ref{fig1}. The first diagram is just the bare vertex.
The second diagram yields up to overall kinematical factors
(Fig.\ref{fig:bare}a)
\begin{equation}
T_{flip}^{2a}=
m_V\frac{g^2}{2\pi} I_1,
\label{eq:t2}
\end{equation}
where
\begin{eqnarray}
I_1 = \int_0^1 \frac{dx}{(1-x)} \frac{ \frac{1}{x}-1}
{p^2-\frac{M^2}{x}-\frac{\lambda^2}{1-x}}
\stackrel{\lambda \rightarrow \infty}{\longrightarrow}
\frac{1}{\lambda^2} \ln \frac{\lambda^2}{M^2}.
\end{eqnarray}
The forth order bare rainbow is a little more complicated, yielding
(up to the same overall factors as Eq. (\ref{eq:t2})
(Fig. \ref{fig:bare}b)
\begin{equation}
T_{flip}^{2b}=
\label{eq:t4}
g^4m_V \int_0^1 \frac{dx/({1-x})}
{P^2-\frac{M^2}{x}-\frac{\lambda^2}{1-x}} \frac{1}{x}
\int_0^x \frac{dy/({x-y})\left(\frac{1}{y}-\frac{1}{x}\right)}
{p^2-\frac{M^2}{y}-\frac{\lambda^2}{x-y}-\frac{\lambda^2}{1-x}}
.
\end{equation}
The $x$ integral  in Eq. (\ref{eq:t4}) is
dominated by small values of $x$ as $\lambda \rightarrow \infty$.
This allows us to simplify the $y$ integral ($z=y/x$)
\begin{eqnarray}
\int_0^x \frac{dy}{x-y} \frac{\left(\frac{1}{y}-\frac{1}{x}\right)}
{p^2-\frac{M^2}{y}-\frac{\lambda^2}{x-y}-\frac{\lambda^2}{1-x}}
%\!\!\!\!\!\!\!\!\!
%\!\!\!\!\!\!\!\!\!
%\!\!\!\!\!\!\!\!\!
%\!\!\!\!\!\!\!\!\!
%\!\!\!\!\!\!\!\!\!
%& & 
%\\ & &
&=&\int_0^1 \frac{dz}{1-z} \frac{\left(\frac{1}{z}-1\right)}
{x\left(p^2-\frac{\lambda^2}{1-x}\right)-\frac{M^2}{z}-\frac{\lambda^2}{1-z}}
\nonumber\\
\stackrel{x \rightarrow 0}{\longrightarrow}&
-&\int_0^1 \frac{dz}{1-z} \frac{\left(\frac{1}{z}-1\right)}
{\frac{M^2}{z}+\frac{\lambda^2}{1-z}}
\stackrel{\lambda \rightarrow \infty}\longrightarrow -I_1 ,
%\nonumber
\end{eqnarray}
i.e.
\begin{equation}
T_{flip}^{2b}= m_V \left(\frac{g^2}{2\pi}I_1\right)^2.
\end{equation}
It turns out that the bare rainbow diagrams form a geometric series,
yielding (together with the bare vertex)
\begin{equation}
T_{flip}^{bare \ rainbow}=\frac{m_V}{1-\frac{g^2}{2\pi}I_1} =
\frac{m_V}{1-\frac{g^2}{2\pi \lambda^2}\ln \frac{\lambda^2}{M^2}}.
\label{eq:tbare}
\end{equation}
Besides the bare rainbow diagrams one also needs to consider the
nested rainbow diagrams (bottom row of Fig. \ref{fig1}), where
the the direction of the chain of instantaneous interactions does
not always alternate along the fermion line. These can obtained
from the bare rainbow diagrams by replacing each instantaneous
interaction by a chain of instantaneous interactions (Fig. \ref{fig:chain})
leading again to a geometric series as $\lambda \rightarrow \infty$
\footnote{Only for $\lambda \rightarrow \infty$ can one neglect the energy
dependence of these bubbles.}, which can be incorporated into above
result by making the replacement
\begin{equation}
g^2 \longrightarrow \frac{g^2}{1- \frac{g^2}{2\pi}I_2},
\label{eq:geo2}
\end{equation}
where
\begin{equation}
I_2 = \int_0^1 \frac{dx}{(1-x)} \frac{1}
{p^2-\frac{M^2}{x}-\frac{\lambda^2}{1-x}}
\stackrel{\lambda \rightarrow \infty}{\longrightarrow}
-\frac{1}{\lambda^2}
\end{equation}
in each instantaneous interaction. Together with Eq. (\ref{eq:tbare}),
this yields for the sum of all planar helicity flip diagrams in leading
order in the vertex mass
\begin{equation}
T_{flip}^{full \ rainbow}=\frac{m_V}{1-\frac{g^2}{2\pi}\frac{I_1} 
{1- \frac{g^2}{2\pi}I_2}} .
\label{eq:tfull}
\end{equation}
However, since 
$I_2/I_1 \sim 1/\ln \lambda^2 
\stackrel{\lambda \rightarrow \infty}{\longrightarrow}0$, one can neglect
$I_2$ in Eq. (\ref{eq:tfull}), yielding
\begin{equation}
T_{flip}^{full \ rainbow} \
\stackrel{\lambda \rightarrow \infty}{\longrightarrow}
\frac{m_V}{1-\frac{g^2}{2\pi \lambda^2}\ln \frac{\lambda^2}{M^2}},
\label{eq:tflip}
\end{equation}
i.e. just the bare result above (\ref{eq:tbare}).

The crucial point is that the denominator of Eq. (\ref{eq:tflip}) becomes 
small in the chiral limit, as one can read off from the gap equation 
(\ref{eq:gap}). In other words, even though each individual diagram
in Fig. (\ref{fig1}) vanishes in the chiral limit $\propto m$, 
summing over an infinite number of diagrams yields 
$%\begin{equation}
T_{flip} \propto M,
$%\end{equation}
which remains finite as $m\rightarrow 0$.

Several interesting observations can be made from this example:
\begin{itemize}
\item While integrating out
zero-modes contribute significantly to the kinetic mass
of fermions, there is no such contribution to the vertex mass from
strict zero-modes.
However, the vertex mass gets renormalized by (infinitesimally)
small $x$ contributions since one
has to sum over an infinite number of Fock space components 
in order to obtain a finite helicity flip amplitude in the chiral limit.
\item In realistic non-perturbative calculations of hadron spectra, where one
cannot include an infinite number of Fock space components, it will at some
level be necessary to absorb the higher order corrections into an effective
vertex mass $M$.
\item The leading diagrams (top row in Fig. \ref{fig1}) have a relatively
simple structure: higher order diagrams can be successively built by
replacing the bare helicity flip vertex inside a given diagram with
the second order dressed helicity flip vertex. This observation may be
useful for a renormalization group study of the helicity flip interactions,
since a large amplitude is obtained only through an infinite chain of steps
from finite $x$ down to vanishingly small values of $x$.
Qualitatively, this mechanism resembles the chiral symmetry breaking 
mechanism suggested in Ref. \cite{lenny}. 
\end{itemize}

%\section{Theta Vacua and Effective LF-Hamiltonians}
%\footnote{The work in this section has been done in collaboration
%with Koji Harada.}
%A very widespread myth is that zero-modes are essential if one wants 
%to describe the physics associated with nontrivial topological effects.
%As an example let us consider
%
\section{Summary}
LF coordinates play a distinguished role in many high energy scattering
experiments (e.g. DIS) and LF quantization represents the most physical 
approach towards a fundamental theoretical description of such experiments.
Lf quantization also yields the very interesting but also controversial
result that LF vacua are trivial. The apparent contradiction between
non-trivial vacua in an equal time formulation and trivial vacua on
the LF is resolved by introducing effective LF Hamiltonians, where
the non-trivial vacuum structure is not ``gone'' but has been absorbed
into effective interaction terms. In a sense, the non-trivial vacuum
structure has been shifted from the states to the operators.
This result was illustrated by studying $QCD_{1+1}$, scalar fields
(1+1 and 3+1 dimensions), Yukawa theories and a (3+1 dimensional) model
with spontaneous breakdown of chiral symmetry.
Finally, I illustrated in a model how a large helicity flip amplitude 
(necessary to produce $\pi-\rho$ or $N-\Delta$ splitting!) emerges in the 
chiral limit by summing over an infinite number of Fock space components.
In a practical numerical calculation (finite number of Fock components), 
this physics must be put in by hand, i.e. by modifying the Hamiltonian.
In the model, this could be accomplished by keeping the vertex
mass finite in the chiral limit (dynamical vertex mass generation).

In these lectures, I have presented a number of examples which show that
effective LF Hamiltonians, with zero modes integrated out, can indeed give
the same physics (e.g. spectrum) as equal time Hamiltonians --- despite the
fact that the LF-vacuum is trivial, and even in situations where symmetries
are spontaneously broken in an equal time framework. These results certainly
constitute an important and encouraging step towards formulating LF-QCD.
Nevertheless, what is still missing is an understanding of how
longitudinal gauge fields and the notorious small $k^+$ divergences arising in
LF gauge fit into above picture.

\section*{Acknowledgements}
I would like to thank C.R. Ji and D.P. Min for inviting me to give
lectures at this interesting summer school.
This work was supported by the D.O.E. under contract DE-FG03-96ER40965
and in part by TJNAF.

%section*{Appendix}

\end{document}